\documentclass[preprint,showpacs,preprintnumbers,amsmath,amssymb]{revtex4}

\usepackage{amsfonts, amssymb,amsmath}
\usepackage{graphicx}
\usepackage{dcolumn}
\usepackage{bm}
\usepackage{subfigure}
\usepackage{epstopdf}
\usepackage{hyperref}
\usepackage{cleveref}

\def \be {\begin{equation}}
\def \ee {\end{equation}}

\begin{document}

\title{Optomechanical cavity without a Stokes side-band}

\author{Devrim Tarhan}
\affiliation{Zafertepe Mah. Kurtulus Cad. No:23 E / 6, K\"{u}tahya, Turkey}  \email{dtarhan@gmail.com}

\date{\today}
\begin{abstract}
We investigate a theoretical demonstration of perfect frequency
conversion in an optomechanical system in the weak coupling regime
without a Stokes side-band. An optomechanical cavity illuminated
by a weak probe field generates two side-modes, differentiating
from the original signal by a phonon frequency. We report the presence of a special combination of parameters in the weak-coupling regime, where
Stokes side-mode vanishes exactly. Only the anti-Stokes mode is
observed with a few hundreds Hz spectral bandwidth of the probe
field. Emergence of this special point is totally unrelated with
the electromagnetically induced transparency (EIT) condition,
where absorption (dip) cancellation is limited with the damping
rate of the mechanical oscillator. Emergence is independent of the
cavity type, i.e. single or double-sided, and takes place only for
a single value of the effective coupling strength constant which
is specific to the system parameters. At a specific effective coupling strength between the mirror and the cavity field, which can be tunable via the coupling field, only the anti-Stokes band is generated. At that specific coupling there appears no Stokes field. Hence, a filter, to eliminate the Stokes field, does not necessitate.
\end{abstract}
\pacs{42.50.Ct, 42.50.Wk} \maketitle
%
%
%

The field of cavity optomechanics investigates the coupling
between light inside an optical cavity and a mechanical oscillator
via radiation pressure force~\cite{aspelmeyer2014cavity}. This
coupling leads to alter light from one cavity resonance to another
resonance of a different frequency in the strong coupling regime
without loss. Optomechanically induced transparency in a membrane for optomechanical cavity setup system was observed experimentally \cite{karuza2013optomechanically}.The optical properties of the optomechanical
system~\cite{oit,agarwal1} controlled by the motion of the
mechanical oscillator in a cavity revealed challenging
applications~\cite{infprocessing1,infprocessing2,infprocessing3,infprocessing4},
especially optomechanical frequency
conversion~\cite{conversion1,conversion2,conversion3,conversion4}.
The conversion between optical fields by using photon-phonon
translation in a cavity optomechanical system has been observed
recently~\cite{conversion1}. And also a wavelength conversion of
light in silicon nitride microdisk resonators has been proposed
experimentally~\cite{conversion2}. In addition to optical
conversion, efficient conversion between microwave and optical
light has been demonstrated reversibly and coherently in the
classical regime~\cite{conversion3}. Finally, microwave frequency
conversion via the motion of a mechanical resonator has been
observed experimentally in the quantum regime~\cite{conversion4}.
A optical frequency conversion can be used in critical functions
such as a quantum network~\cite{kimble}. Cavity optomechanics
systems allow for fundamentally new forms of quantum
information~\cite{quantuminfo1} and optomechanical light
storage~\cite{lightstorage}. Since not only photons are
non-interacting particles but also insensitive to environmental
distortion, photons are preferred for optical communication and
quantum information processing which requires interaction. The
interaction between light inside an optical cavity and a
mechanical oscillator can be used for this procedure.
Optomechanical systems have become attractive for quantum
optomechanical memory~\cite{painter} and quantum state
transfer~\cite{statetransfer1,statetransfer2} very recently.

The coupling between light inside an optical cavity and a
mechanical oscillator can be used in telecommunications for signal
modulation since this type of interaction leads to change the
frequency of the incident beam and its
wavelength~\cite{conversion1}. In this work, we investigate the
frequency conversion via an optomechanical system operating in the
weak coupling regime without a Stokes side-band. In an
optomechanical cavity system, mechanical frequency (nanomechanical
oscillator of frequency $\omega_m$) will modulate the probe laser
with frequency $\omega_p$ and will form  a new sidebands frequency
$2\omega_{L}-\omega_{p}$, which is called anti-Stokes sidebands.
We find that the conversion can be controlled by the
optomechanical coupling strength and detuning. We report the
presence of a special combination of parameters in the
weak-coupling regime($g \ll \kappa,\gamma$), where Stokes
side-mode vanishes exactly. Only the anti-Stokes mode is observed
with a few hundreds Hz spectral bandwidth. Emergence of this
special point is totally unrelated with the EIT condition, where
absorption cancellation is limited with the damping rate of the
mechanical oscillator. Emergence is independent of the cavity
type, i.e. single or double-sided, and takes place only for a
single value of the effective coupling constant which is specific
to the system parameters. We report a theoretical demonstration of
the frequency conversion in an optomechanical system in the weak
coupling regime. Beside the microwave frequency conversion by the motion of a mechanical resonator ~\cite{conversion4} our system show that by tuning the probe field at a special interaction strength $g$ the classical optical field can be inverted without loss.

Firstly, we briefly describe the Hamiltonian of the systems. We
obtain the Heisenberg equations and then we get the first order
solutions analytically. An optomechanical system consists of an
optical cavity and  a nanomechanical mirror which can be treated
as nano mechanical oscillator(NMO). NMO is located at the center
of the cavity. The cavity field (frequency $\omega_c$) is driven
by an external strong source with frequency $\omega_L$ and a weak
probe field of frequency $\omega_p$.

\begin{figure}[!t]
\begin{center}
\includegraphics[width=8.3cm]{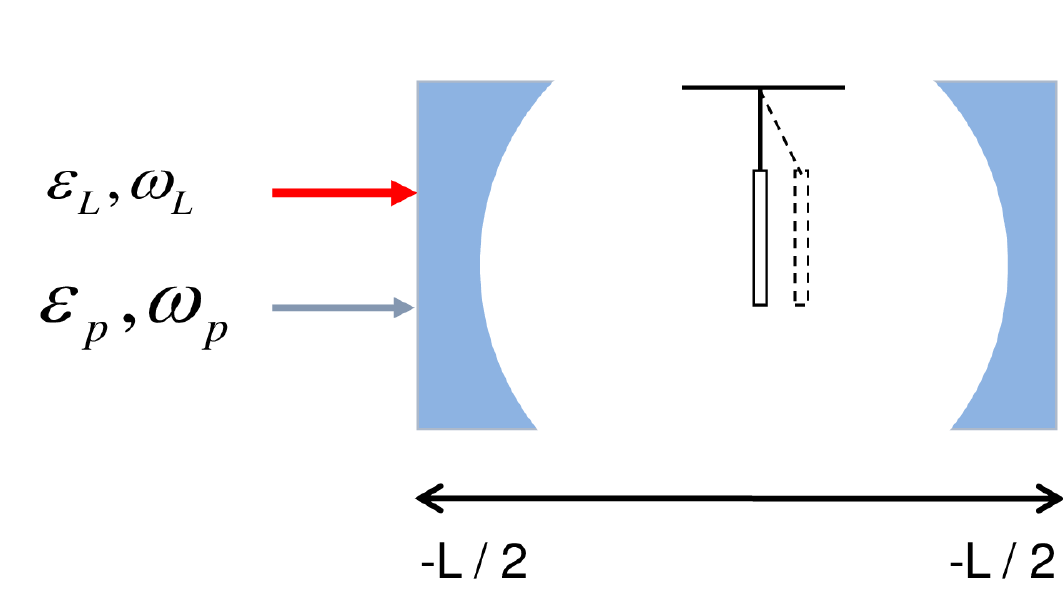}
\caption{\label{fig1}(Color online) Schematic of a double-ended
cavity with a moving nanomechanical mirror. The length of the
cavity is $L$. The classical probe field $\varepsilon_p$ and the
coupling field $\varepsilon_L$ are applied to the cavity.}
\end{center}
\end{figure}

The probe field $\varepsilon_p$ is treated as classical and the
response of the cavity optomechanical system to the probe field in
the presence of the coupling field $\varepsilon_L$ is computed.
Mechanical-cavity field coupling is provided by presence of an
coupling laser. The mass and resonance frequency of an
nanomechanical oscillator is $m$ and $\omega_m$ respectively. NMO
is coupled to a Fabry-Perot cavity via radiation pressure effects
~\cite{aspelmeyer2014cavity}. In a Fabry-Perot cavity, both
mirrors have equal reflectivity. We use a configuration in which a
partially transparent NMO is in the middle of a cavity that is
bounded by two high-quality mirrors as shown in Fig. ~\ref{fig1}.
%
The Hamiltonian of this system is given by ~\cite{tarhan}

\begin{eqnarray} \label{Ham}
H  &=& \hbar \Delta_c  c^{\dagger} c - \hbar g_0 c^{\dagger} c q +
\frac{p^2}{2m}+\frac{1}{2}m \omega_m^2 q^2 \nonumber \\ &+& i\hbar
\varepsilon_L (c^{\dagger} - c ) + i\hbar
(c^{\dagger}\varepsilon_p e^{-i\Delta_{p} t}-c \varepsilon_p^\ast
e^{i\Delta_{p} t}).
\end{eqnarray}
$\Delta_{c}=\omega_c-\omega_L$, $\Delta_{p}=\omega_p-\omega_L$ are
the frequencies of cavity and probe field. $g_0=\omega_c / L$ is
the coupling constant between the cavity field and the movable
mirror ~\cite{meystre1}, $q_0$ is the steady state position of the
movable mirror and $c,c^{\dagger}$ are the annihilation creation
and  operators of the photons of the cavity field respectively.
The momentum and position operators of the nanomechanical
oscillator are $p$ and $q$, respectively. $\varepsilon_L$ is taken
to be real while $\varepsilon_p$ is taken to be complex. The
amplitude of the pump field is $\varepsilon_L=\sqrt{2\kappa
P_L/\hbar \omega_L}$ with $P_L$ being the pump power.

Heisenberg equation of motion for the coupled cavity-mirror system
is written and the damping rate  $2\kappa$ is added
phenomenologically  to represent the loss at the cavity mirrors
where $\kappa$ is the decay rates of the cavity mirrors. The
damping rate of the mechanical oscillator is $\gamma$. Here we
neglect the quantum and thermal noise and examine the system in
the mean field limit ~\cite{agarwal1,tarhan}. The linear response solution was developed analytically: ~\cite{tarhan}
\begin{equation}
\label{linearsol}
c(t) = c_{0} + c_{+} \varepsilon_p e^{-i \Delta_{p} t} + c_{-} \varepsilon_p^{\star} e^{i \Delta_{p} t}.
\end{equation}
If we solve
Heisenberg equation of motion by using the linear response
solution \cite{boyd,tarhan}, we
will get the steady state solutions $c_0 = \varepsilon_L/(2
\kappa+i \Delta)$, $ q_0 = - \hbar g |c_0|^2/m \omega_m^2$ and the
first order solutions $c_{+}$ and $c_{-}$ ~\cite{tarhan}:
\begin{eqnarray} \label{cplus}
c_{+} &=& \frac{A\times B-i \omega_m
|g|^2}{[(\kappa-i\Delta_{p})^2+\Delta^2] \times B+2 \omega_m
\Delta |g|^2 }
\nonumber \\
 c_{-}
&=& \frac{i |g|^2
\omega_m}{[(\kappa-i\Delta_{p})^2+\Delta^2]\times B +2 \omega_m
\Delta |g|^2 },
\end{eqnarray}
where $\Delta=\Delta_c - g_0q_0 $ is the effective detuning,
$A=[\kappa-i(\Delta+\Delta_{p})]$, $B=(\Delta_p^2-\omega_m^2+i
\gamma_m \Delta_{p})$, $g=g_0c_0$ is the effective coupling,
$|c_0|^2$ is the resonator intensity which is obtained from steady
state equations, and $q_0$ is the steady state position of the
movable mirror.


 The output field can be obtained by using the
input-output relation $\varepsilon_{out}=2\kappa \langle c \rangle
$~\cite{sumei}. According to linear solution written in Eq. (\ref{linearsol}) , we expand the output field to the first order in the probe
field $\varepsilon_p$ and find the output field as
\begin{equation}
\label{outputfiledlinearsol}
\varepsilon_{out}(t) = \varepsilon_{out0} + \varepsilon_{out+}
\varepsilon_p e^{-i \Delta_{p} t} + \varepsilon_{out0-}
\varepsilon_p^* e^{i \Delta_{p} t},
\end{equation}
where $\varepsilon_{out0}$, $\varepsilon_{out+}$, and
$\varepsilon_{out-}$ are the components of the output field
oscillating at frequencies  $\omega_{L}$, $\omega_{p}$, and
$2\omega_{L}-\omega_{p}$. The terms $e^{\pm i 2 \Delta_{p} t}$ are
not taken into account because we are interested in the linear
response Refs.~\cite{agarwal1,tarhan,boyd}.

\begin{figure}[!t]
\centering{\vspace{0.5cm}}
\begin{center}
\includegraphics[width=15cm]{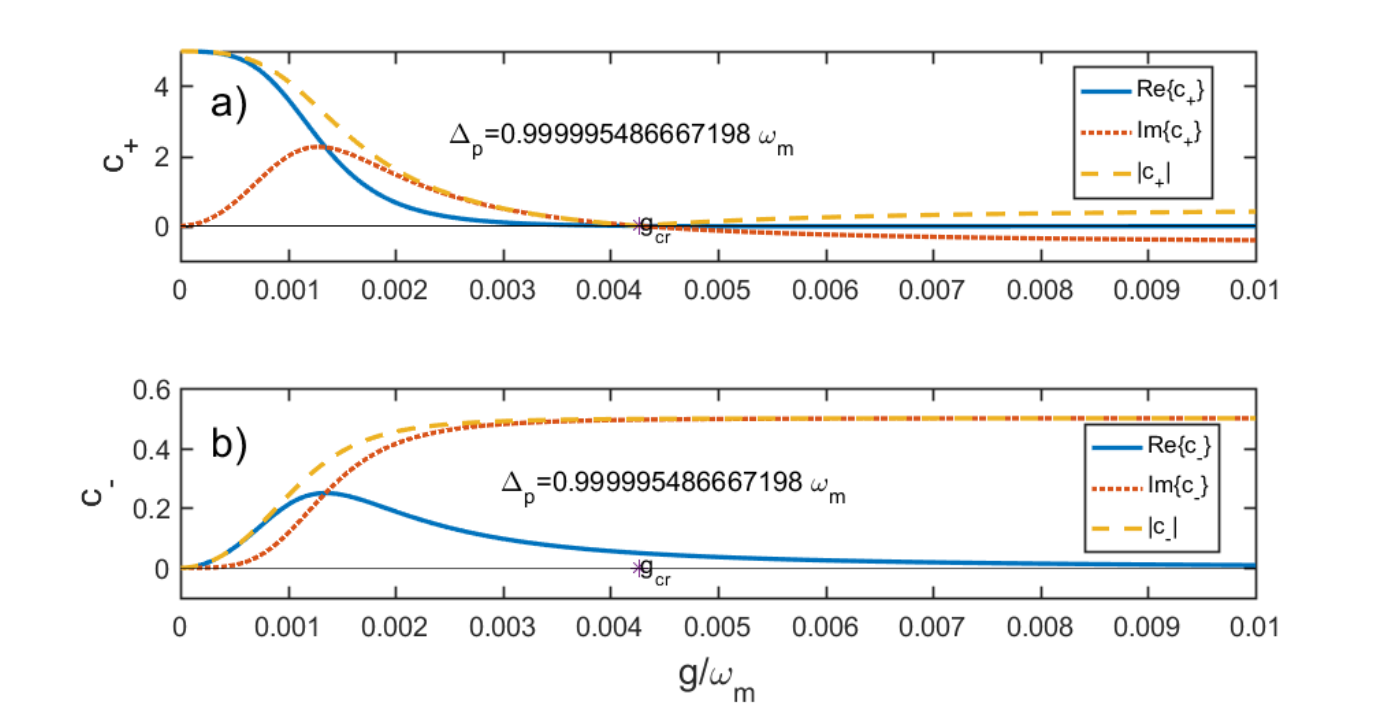}
\caption{\label{fig2} The analytical solution $c_{+}$ and $c_{+}$
as a function of the optomechanical coupling strength. The
parameters used are the length of the cavity $L=6.7$ cm, the
wavelength of the laser $\lambda=2\pi c/\omega_c=1064$ nm, $m=40$
ng, $\omega_m=2\pi \times 134$ kHz, $\gamma=0.76$ Hz,
$\kappa=\omega_m/10$ and $\Delta_p=0.999995486667198 \omega_m$,
$\Delta=\omega_m$ and $g_{cr}=0.0043\omega_m$.}
\end{center}
\end{figure}

In our calculations, we use experimental parameters
\cite{thompson} for the length of the cavity $L=6.7$ cm, the
wavelength of the coupling laser $\lambda=2\pi c/\omega_c=1064$
nm, $m=40$ ng, $\omega_m=2\pi \times 134$ kHz, $\gamma=0.76$ Hz,
$\kappa=\omega_m/10$, and $\Delta=\omega_m$. Analytical
demonstration of the first order solutions are given in Fig.
~\ref{fig2}. We show the real and the imaginary parts of the
$c_{+}$ and $c_{-}$ In Fig. ~\ref{fig2}. In order to cancel the
Stokes field, we equalize the absolute value of numerator $c_{+}$
to the zero. After solving the cubic equation, we get only one
real value $\Delta_p=0.999995486667198 \omega_m$ at critical
coupling strength $g=0.0043 \omega_m$. This special point is
totally unrelated with the EIT condition, where absorption
cancellation is limited with the damping rate of the mechanical
oscillator.
The value of imaginary and real parts $c_{+}$ are both 0 when the
probe detuning at a special point $\Delta_p=0.999995486667198 \omega_m$. It has been easily seen in Fig. ~\ref{fig2} that
imaginary and real parts of $c_{+}$ are zero in the weak-coupling
regime $g=0.0043\omega_m$. One can infer from Fig. ~\ref{fig2}
that the component of the output field oscillating at frequency
$\omega_{p}$ disappears whereas the component of the output field
oscillating at frequency $2\omega_{L}-\omega_{p}$ will occur.
\begin{figure}[!t]
\centering{\vspace{0.5cm}}
\begin{center}
\includegraphics[width=9cm]{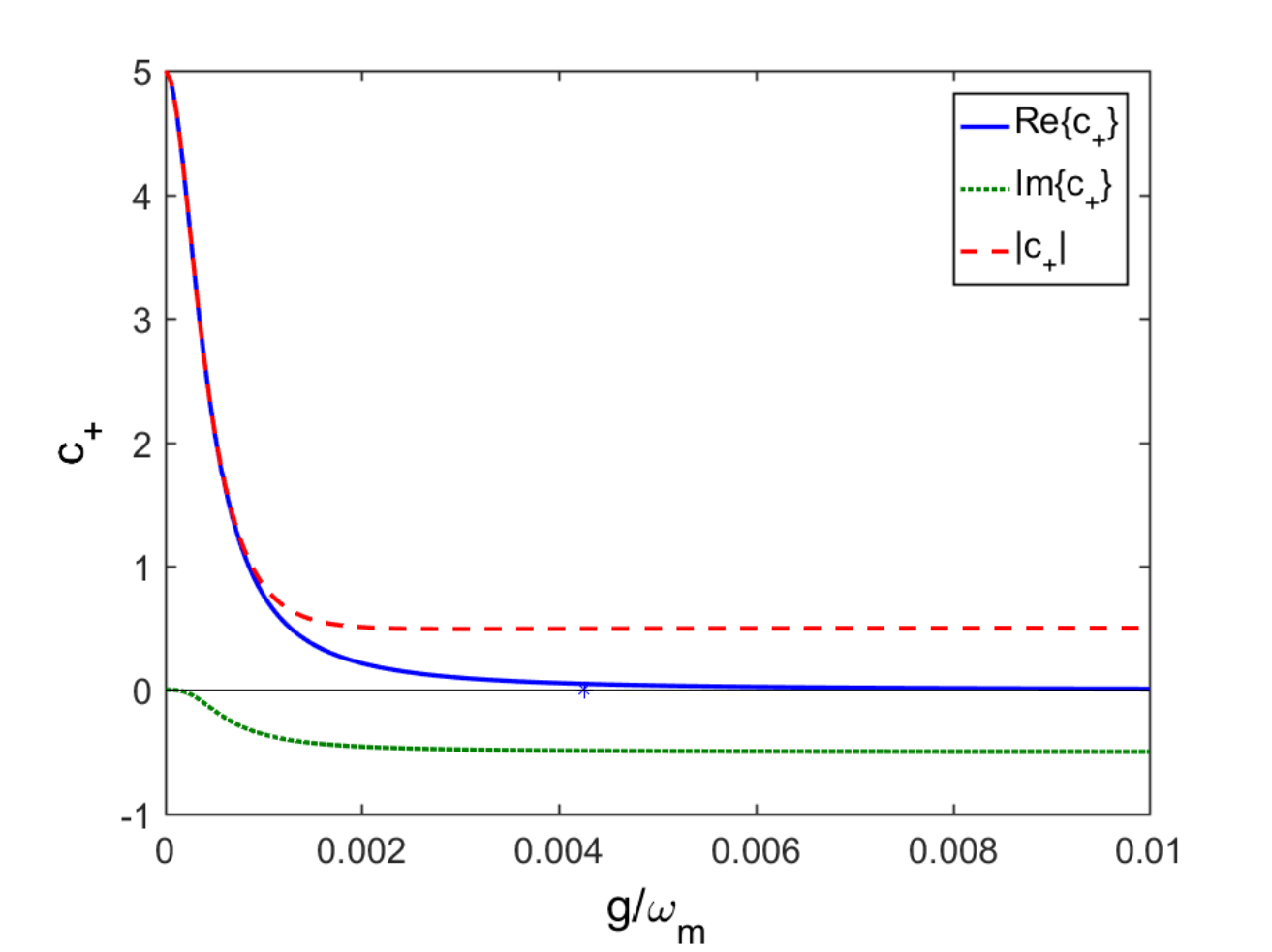}
\caption{\label{fig3} The analytical solution $c_{+}$. The
parameters are the same as in Fig. \ref{fig2}, except that
$\Delta_p=\omega_m$ (EIT condition).}
\end{center}
\end{figure}
%
\begin{figure}[htb]
\centering \subfigure[]{
\includegraphics[width=0.4\textwidth]{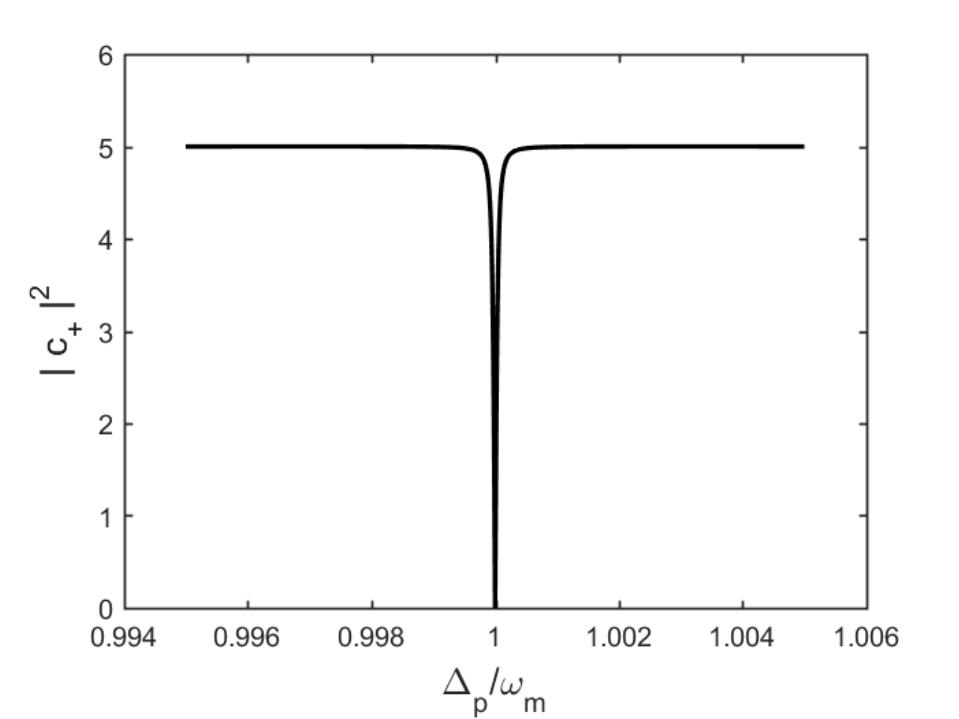}
\label{fig:4a} } \subfigure[]{
\includegraphics[width=0.4\textwidth]{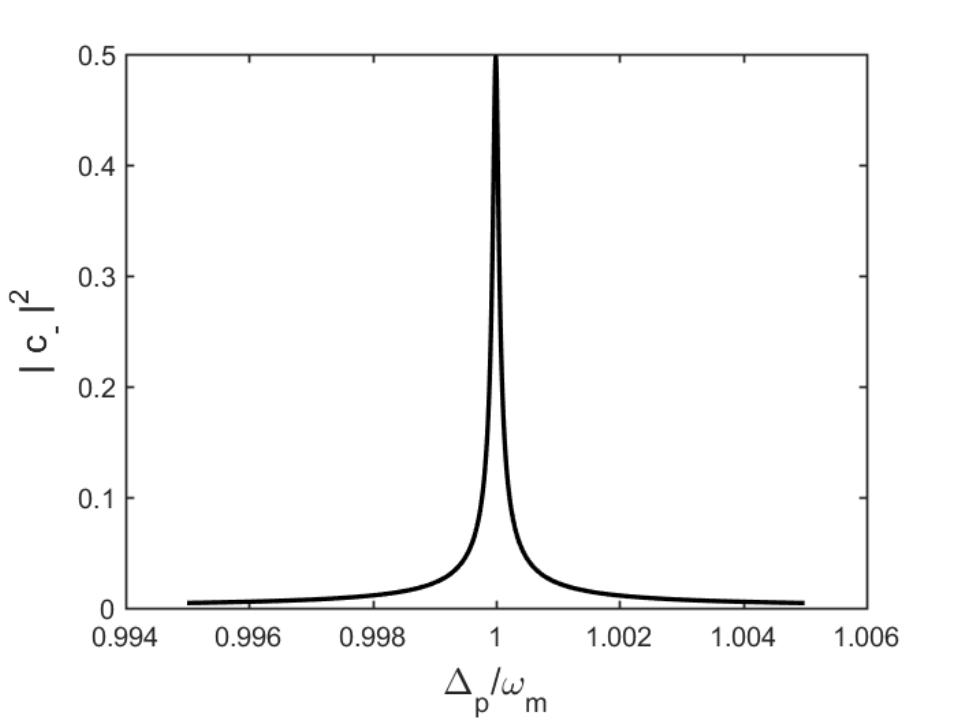}
\label{fig:4b} } \caption{(a) The intensity $|c_{+}|^2$ as a
function of the probe detuning $\Delta_p$. (b) The intensity
$|c_{-}|^2$ as a function of the probe detuning $\Delta_p$. The
parameters used for both curves are the same as in Fig.
\ref{fig2}.} \label{fig4}
\end{figure}

At the same time we plot the $c_{+}$ as a function of $g$ in Fig.
\ref{fig3} when the probe detuning $\Delta_p=\omega_m$ in order to
show the importance of the special point. As seen in in Fig.
~\ref{fig3} the value of imaginary part of $c_{+}$ is different
from 0 at the probe detuning $\Delta_p=\omega_m$.

Moreover, we plot the intensity spectrums $|c_{+}|^2$ and
$|c_{-}|^2$ as a function of the probe detuning $\Delta_p$ in Fig.
\ref{fig4}. As seen in Fig. \ref{fig4} a few hundreds Hz spectral
bandwidth of the probe field(only the anti-Stokes mode) is
observed. The value of $|c_{+}|^2$ is completely is 0 whereas the
value of $|c_{-}|^2$ is about 0.25 when the probe detuning
$\Delta_p=0.999995486667198 \omega_m$ at a critical coupling strength
$g=0.0043\omega_m$. The intensity of the $|c_{+}|^2$ which is
called linear response vanishes at these special points. At these
points the linear response disappears and anti-Stokes field occurs
which is shown in Figs. ~\ref{fig2} and ~\ref{fig4}. In the
presence of a special combination of parameters in the
weak-coupling regime, where Stokes side-mode vanishes exactly.
Only the anti-Stokes mode is observed with a few hundreds Hz
spectral bandwidth.

We have reported a theoretical demonstration of perfect frequency
conversion in an optomechanical system in the weak coupling regime
without a Stokes side-band. We have converted the classical
optical fields with frequency $\omega_p$ to another optical fields
with frequency $2 \omega_L-\omega_p$. We report the presence of a
special combination of parameters ---in the weak-coupling regime,
e.g. $g=0.0043 \omega_m$--- where Stokes side-mode vanishes
exactly. This property is clearly beneficial for demonstration of
frequency conversion in an optomechanical setup which is formed by
vibrating mirror under the action of a strong coupling field. It
has been shown that, in the a few hundreds Hz spectral width of
the probe field, only the anti-Stokes mode is observed.

Furthermore we have computed the intensity spectrum of the probe
field. Our results reveal that by tuning the probe field at a
special interaction strength $g$ the classical optical field can
be inverted without loss. This result can be useful for quantum
information processing because there is no signal loss in our
system.
\begin{acknowledgements}
D.T. gratefully thanks to N. Postacioglu for computer support and thanks to O. E. M\"ustecapl{\i}o\u{g}lu for his useful discussions.
\end{acknowledgements}
\bibliography{conversionPRAbib}{}

\end{document}